\documentstyle[amssymb,preprint,aps]{revtex}

\begin{document}
\title{Testing Broken $U(1)$ Symmetry in a Two-Component Atomic Bose-Einstein
Condensate}
\author{C. P. Search and P.R. Berman}
\address{Michigan Center for Theoretical Physics and Physics Department, University\\
of Michigan, Ann Arbor, MI, 48109-1120}
\date{\today}
\maketitle
\pacs{03.75.Fi, 05.30.Jp, 32.80.Pj}

\begin{abstract}
We present a scheme for determining if the quantum state of a small trapped
atomic Bose-Einstein condensate is state with a well defined number of
atoms, a Fock state, or a state with a broken $U(1)$ gauge symmetry, a
coherent state. The proposal is based on the observation of Ramsey fringes.
The population difference measured by a Ramsey fringe experiment will
exhibit collapse and revivals due to the mean-field interactions. The
collapse and revival times depend on the relative strengths of the
mean-field interactions for the two components and the initial quantum state
of the condensate.
\end{abstract}

\section{Introduction}

Since the observation of Bose-Einstein condensation in trapped atomic gases
in 1995 \cite{BEC}, an unresolved issue in the theory of atomic
Bose-Einstein condensates (BEC) is the qauntum state of the condensate \cite
{barnett}\cite{wright1}\cite{jav}\cite{wright2}\cite{collett}\cite
{lewenstein}\cite{imamoglu}\cite{dunning}. In the standard theory of BEC,
which applies in the thermodynamic limit, the quantum state is one of well
defined phase, $\phi $. This state corresponds to a broken $U(1)$ gauge
symmetry \cite{Anderson}\cite{leggett}. Because particle number, $N$, and
phase obey the uncertainty relation $\Delta N\Delta \phi \geq 1$ \cite{Huang}%
, a state of well defined phase implies uncertainty in the particle number.

Again, this result is based on assumptions which are valid only in the
thermodynamic limit in which $N\rightarrow \infty .$ The second quantized
Hamiltonian for a system of bosons is expressed in terms of the bosonic
field operator, $\hat{\Psi}({\bf r},t{\bf ),}$ and its adjoint which
annihilate and create a particle at position ${\bf r,}$ respectively. The
Hamiltonian is invariant under the global $U(1)$ transformation, $\hat{\Psi}(%
{\bf r},t{\bf )}\rightarrow e^{i\chi }\hat{\Psi}({\bf r},t{\bf ).}$ The $%
U(1) $ symmetry implies a conserved Noether charge which corresponds to the
total number of particles. When a Bose-Einstein condensate is present, a
single quantum state becomes macroscopically occupied and it is assumed that
the field operator acquires a nonvanishing expectation value,

\begin{equation}
\left\langle \hat{\Psi}({\bf r},t{\bf )}\right\rangle =\psi ({\bf r},t{\bf ),%
}  \label{SBGS}
\end{equation}
with respect to the condensate state. Here, $\psi ({\bf r},t{\bf )}$ is the
order parameter for the condensate. However, $\psi ({\bf r},t{\bf )}$ is no
longer invariant under the transformation, $\psi ({\bf r},t{\bf )}%
\rightarrow e^{i\chi }\psi ({\bf r},t{\bf )}$ which implies that the $U(1)$
symmetry is spontaneously broken in the condensate. Equation (\ref{SBGS})
may be rewritten as $\left\langle \hat{\Psi}({\bf r},t{\bf )}\right\rangle
=\left\langle \varphi (N-1)\right| \hat{\Psi}({\bf r},t{\bf )}\left| \varphi
(N)\right\rangle $ where $\left| \varphi (N)\right\rangle $ and $\left|
\varphi (N-1)\right\rangle $ are ''like'' condensate states which differ by
one particle \cite{landau}\cite{pines}. In the thermodynamic limit, the
difference between $\left| \varphi (N)\right\rangle $ and $\left| \varphi
(N-1)\right\rangle $ disappears, in which case the condensate is in a
coherent state, $\hat{\Psi}({\bf r},t{\bf )}\left| \varphi (N)\right\rangle
=\psi ({\bf r},t{\bf )}\left| \varphi (N)\right\rangle $ and $\psi ({\bf r},t%
{\bf )}$ may be identified with the wave function for the quantum state in
which Bose condensation has occurred (with the wave function normalization $%
\int d^{3}r\left| \psi ({\bf r},t{\bf )}\right| ^{2}=N$). However, it is not
clear that Eq. (\ref{SBGS}) is still applicable when $N$ is finite. Examples
of BEC in condensed matter physics such as superfluid He may have $N\sim
10^{20}$ whereas trapped atomic gases typically have $N\sim 10^{3}-10^{6}.$

In fact, there are two immediate objections to the use of a coherent state
for finite particle number. First, at zero temperature, one expects that the 
{\it true} ground state of the condensate will be a number state (i.e. a
Fock state) with no {\it quantum} fluctuations in the particle number even
if we are ignorant of what that number is. Also, since the Hamiltonian for
the system is $U(1)$ symmetric, one must introduce a symmetry breaking field
into the Hamiltonian in order to define the phase of the condensate \cite
{bogo}. The symmetry breaking field vanishes in the thermodynamic limit.
However, there is no physical interaction which corresponds to this symmetry
breaking term and as such, it simply amounts to a mathematical trick \cite
{jav2}.

The general definition of a Bose-Einstein condensate due to Penrose and
Onsager \cite{ODLRO} is that the single particle density matrix, $\rho _{1}(%
{\bf r},{\bf r}^{\prime },t)=\left\langle \hat{\Psi}^{\dagger }({\bf r},t%
{\bf )}\hat{\Psi}({\bf r}^{\prime },t{\bf )}\right\rangle $, does not vanish
as for large separations, 
\begin{equation}
\lim_{|{\bf r}-{\bf r}^{\prime }|\rightarrow \infty }\rho _{1}({\bf r},{\bf r%
}^{\prime },t)=\Phi ^{\ast }({\bf r},t{\bf )}\Phi ({\bf r}^{\prime },t{\bf ).%
}  \label{odlro}
\end{equation}
Although Eq. (\ref{odlro}) is consistent with Eq. (\ref{SBGS}), Eq. (\ref
{odlro}) simply requires the macroscopic occupation of a single quantum
state and will therefore be true for a condensate that is in a number state.
Wright {\it et al. }have shown that, for small condensates ($N\sim 10^{3}$)
described initially by a coherent state, the order parameter undergoes
collapses and revivals such that $\psi ({\bf r},t{\bf )}\rightarrow 0$
during the collapse but Eq. (\ref{odlro}) remains valid at all times since $%
\rho _{1}({\bf r},{\bf r}^{\prime },t)$ is unaffected by the phase diffusion
which causes the collapse and revivals \cite{wright1}. This implies that a
coherent state description is inappropriate for small $N$ since it is not an
energy eigenstate of the system. Similar results were obtained in \cite
{imamoglu} where the depedence of the collapse and revival times on the
dimensionality of the condensate and the trapping potential was studied. In
contrast, Barnett {\it et al. }have argued that the best pure state
description of a condensate is the coherent state since it is the most
robust state with respect to interactions with the environment \cite{barnett}%
. In short, there appear to be no conclusive arguments for or against a
coherent state description of atomic BEC's.

Since a coherent state has a well defined phase, the appearance of
interference fringes in the atomic density for two overlapping condensates
would be an indication that the condensates were in coherent states (or some
other superposition of number states such that one could ascribe a phase to
the condensate). However, Javanainen and Yoo have shown that even if the two
condensate states are initially in number states, there will be an
observable interference pattern \cite{jav}. This is because the destructive
detection of atoms creates an uncertainty in the relative number of atoms in
the two condensates since it not known from which condensate the detected
atom came from. Consequently, with each atom detection, the relative phase
between the two condensates becomes more precisely defined. Thus, any
interference experiment based on destructive detection of atoms will not be
able to distinguish between two condensates initially in number states or
coherent states. Similar work has shown that the detection of spontaneously
scattered photons between two condensates can establish a relative phase
between the condensates even when the condensates are initially in number
states \cite{ruos}.

In this paper, we propose a method for distinguishing between a condensate
that is in a number state and a coherent state. The method is based Ramsey's
separated oscillatory field technique \cite{ramsey} in a two-component
condensate such as $^{87}Rb$ \cite{Rb}. Such an experiment has recently been
performed by Hall {\it et al. }at JILA \cite{relphase}. Ramsey's technique,
as applied to two-level atoms initially in their ground states, consists of
applying two ''$\frac{\pi }{2}$-pulses'' generated by an external field of
frequency $\omega _{e}$ which couple the ground state and excited state.
These pulses are separated by a time $T$. The first pulse puts each of the
atoms into a superposition of the ground and excited states with equal
population. The relative phase between the two states then evolves as $%
\omega _{o}T$ where $\hbar \omega _{o}$ is the energy difference between the
two states. The second pulse creates a population difference between the two
states which measures the relative phase accummulated by the atoms as
compared to the phase accummulated by the external field during the period $T
$. The population difference is then $\cos (\delta T)$ where $\delta =\omega
_{o}-\omega _{e}$. In a condensate, the population difference after the
second pulse will be affected by two-body interactions which cause a phase
diffusion of the relative phase of the two components in the interval $T$
between the pulses. As such, the population difference will experience
collapse and revivals as function of $T$. The collapse and revival times
depend on the strength of the two-body interactions and the intial state of
the condensate such that the collapse and revival times for a coherent state
are different from a number state.

Wright {\it et al.} predicted a similar effect for the interference fringe
visibility of two spatially overlapping condensates \cite{wright2}. They
showed that the revival time for the fringe visibility for condensates
initially in coherent states was twice that of condensates in number states.
However, their result was based on the assumption that the intra-condensate
interactions were the same and that inter-condensate interactions could be
ignored. They also assumed that the coherence between the two number state
condensates was established by measurement of the interference pattern in
the same manner as described in \cite{jav}.

The key advantage of the Ramsey fringe technique is that the collapse and
revivals manifest themselves in the population difference between the two
condensate components which is readily measured used absorptive or
dispersive imaging of the condensate. Proposals to directly measure the
order parameter, in order to detect collapse and revivals, such as \cite
{light} usually rely on the detection of scattered light from the
condensate. Reference \cite{light} involves two independent condensates that
are in spatially separated potentials. Such an experiment would be
technically difficult.

The remainder of this paper is organized as follows: In section II we
present the second quantized Hamiltonian for a two-component condensate and
derive a two-mode model for the ground states of the two components. The
two-mode Hamiltonian is then represented in terms of angular momentum
operators by exploiting the equivalence between the algebra of two harmonic
oscillators and the angular momentum algebra. In section III, we consider a
condensate prepared in one of the modes with a state vector given by either
a number state or a coherent state and study the time evolution of these
states subject to two $\frac{\pi }{2}$ pulses. In section IV and V, we
discuss the collapse and revivals as well as relevant time scales for
observing them.

\section{Physical Model}

\subsection{Derivation of two-mode Hamiltonian}

We consider a collection of bosonic atoms that have internal states $\left|
1\right\rangle $ and $\left| 2\right\rangle $ with energies $\hbar \omega
_{o}/2$ and $-\hbar \omega _{o}/2$, respectively. There is a spatially
uniform time dependent radiation field with frequency $\omega _{e}$ which
couples the two internal states with a Rabi frequency $\Omega (t)$. The atom
field detuning is denoted by $\delta =\omega _{o}-\omega _{e}$. The atoms in
states $\left| 1\right\rangle $ and $\left| 2\right\rangle $ are subject to
isotropic harmonic trapping potentials $V_{i}({\bf r)=}\frac{1}{2}m\omega
_{i}r^{2}$ for $i=\{1,2\}$, respectively. Furthermore, the atoms interact
via elastic two-body collisions through the interaction potentials $V_{ij}(%
{\bf r}-{\bf r}^{\prime })=U_{ij}\delta \left( {\bf r-r}^{\prime }\right) $
where $U_{ij}=\frac{4\pi \hbar ^{2}a_{ij}}{m}$ and $a_{ij}$ is the s-wave
scattering length between atoms in states $i$ and $j$. It is assumed that $%
a_{ij}>0$ corresponding to repulsive interactions. The Hamiltonian operator
describing the system is given by, 
\begin{mathletters}
\begin{eqnarray}
\hat{H} &=&\hat{H}_{atom}+\hat{H}_{coll}  \label{ham} \\
\hat{H}_{atom} &=&\int d^{3}r\left\{ \hat{\Psi}_{1}^{\dagger }({\bf r})\left[
-\frac{\hbar ^{2}}{2m}\nabla ^{2}+V_{1}({\bf r)+}\frac{\hbar \delta }{2}%
\right] \hat{\Psi}_{1}({\bf r})+\hat{\Psi}_{2}^{\dagger }({\bf r})\left[ -%
\frac{\hbar ^{2}}{2m}\nabla ^{2}+V_{2}({\bf r)-}\frac{\hbar \delta }{2}%
\right] \hat{\Psi}_{2}({\bf r})\right.   \nonumber \\
&&\left. +\frac{\hbar }{2}\Omega ^{\ast }(t)\hat{\Psi}_{1}^{\dagger }({\bf r}%
)\hat{\Psi}_{2}({\bf r})+\frac{\hbar }{2}\Omega (t)\hat{\Psi}_{2}^{\dagger }(%
{\bf r})\hat{\Psi}_{1}({\bf r})\right\}   \label{hatom} \\
\hat{H}_{coll} &=&\frac{1}{2}\int d^{3}r\left\{ U_{11}\hat{\Psi}%
_{1}^{\dagger }({\bf r})\hat{\Psi}_{1}^{\dagger }({\bf r})\hat{\Psi}_{1}(%
{\bf r})\hat{\Psi}_{1}({\bf r})+U_{22}\hat{\Psi}_{2}^{\dagger }({\bf r})\hat{%
\Psi}_{2}^{\dagger }({\bf r})\hat{\Psi}_{2}({\bf r})\hat{\Psi}_{2}({\bf r}%
)\right.   \nonumber \\
&&\left. +2U_{12}\hat{\Psi}_{1}^{\dagger }({\bf r})\hat{\Psi}_{2}^{\dagger }(%
{\bf r})\hat{\Psi}_{1}({\bf r})\hat{\Psi}_{2}({\bf r})\right\} .
\label{hcoll}
\end{eqnarray}
Here, $\hat{H}_{atom}$ is the single particle Hamiltonian and $\hat{H}_{coll}
$ represents two-body interactions.

The operators $\hat{\Psi}_{i}({\bf r})$ and $\hat{\Psi}_{i}^{\dagger }({\bf r%
})$ are bosonic annihilation and creation operators for an atom in state $%
i=\{1,2\}$ at position ${\bf r}$ which satisfy the commutation relations $[%
\hat{\Psi}_{i}({\bf r}),\hat{\Psi}_{j}^{\dagger }({\bf r}^{\prime })]=\delta
_{ij}\delta \left( {\bf r-r}^{\prime }\right) $ and $[\hat{\Psi}_{i}({\bf r}%
),\hat{\Psi}_{j}({\bf r}^{\prime })]=0$. The operators, $\hat{\Psi}_{i}({\bf %
r}),$ have been written in a field interaction representation which is
rotating at the frequency of the external field, $\omega _{e}$, so that $%
\hat{\Psi}_{1}({\bf r})=\hat{\Psi}_{1}^{(N)}({\bf r})e^{i\omega _{e}t/2}$
and $\hat{\Psi}_{2}({\bf r})=\hat{\Psi}_{2}^{(N)}({\bf r})e^{-i\omega
_{e}t/2}$. Here, $\hat{\Psi}_{i}^{(N)}({\bf r})$ are the field operators in
the normal representation. This explains the appearance of the detuning in
Eq. (\ref{hatom}).

In the presence of the condensate, we assume that the field operators may be
approximated using a two-mode model such that $\hat{\Psi}_{1}({\bf r}%
)=a_{1}\phi _{1}({\bf r})+\delta \hat{\Psi}_{1}({\bf r})$ and $\hat{\Psi}%
_{2}({\bf r})=a_{2}\phi _{2}({\bf r})+\delta \hat{\Psi}_{2}({\bf r})$ where
the $a_{i}$ are the mode annihilation operators for the condensate modes
which obey bosonic commutation relations $\left[ a_{i},a_{j}^{\dagger }%
\right] =\delta _{ij}$ and $\left[ a_{i},a_{j}\right] =0.$ The $\delta \hat{%
\Psi}_{i}({\bf r})$ represent the field operator for the non-condensate
modes and will be neglected since the number of atoms in these modes is
assumed to be negligible compared to the condensate modes. For small
condensates, such that $Na_{ij}/a_{ho,i}\lesssim 1$ where $a_{ho,i}=\sqrt{%
\frac{\hbar }{m\omega _{i}}}$is the harmonic oscillator length, $\phi _{i}(%
{\bf r})$ are given by the harmonic oscillator ground states of the trap, $%
\left[ -\frac{\hbar ^{2}}{2m}\nabla ^{2}+V_{i}({\bf r)}\right] \phi _{i}(%
{\bf r})=\frac{3\hbar \omega _{i}}{2}\phi _{i}({\bf r})$ \cite{mode}.
Assuming a weak trap, $a_{ho,i}\approx 10\mu m$ and $a_{ij}\approx 5nm$ for $%
^{87}Rb$, one has $N\lesssim 2000.$ In the two-mode approximation the
Hamiltonian becomes (with $\hbar =1$), 
\end{mathletters}
\begin{eqnarray}
\hat{H} &=&\frac{1}{2}\left( \delta +3\omega _{1}\right) a_{1}^{\dagger
}a_{1}+\frac{1}{2}\left( -\delta +3\omega _{2}\right) a_{2}^{\dagger }a_{2}+%
\frac{1}{2}\bar{\Omega}^{\ast }(t)a_{1}^{\dagger }a_{2}+\frac{1}{2}\bar{%
\Omega}(t)a_{2}^{\dagger }a_{1}  \nonumber \\
&&+\frac{1}{2}\left( \chi _{1}a_{1}^{\dagger }a_{1}^{\dagger
}a_{1}a_{1}+\chi _{2}a_{2}^{\dagger }a_{2}^{\dagger }a_{2}a_{2}+2\chi
_{12}a_{1}^{\dagger }a_{1}a_{2}^{\dagger }a_{2}\right) ;  \label{mode}
\end{eqnarray}
where 
\begin{mathletters}
\begin{eqnarray}
\bar{\Omega}(t) &=&\Omega (t)\int d^{3}r\phi _{2}^{\ast }({\bf r})\phi _{1}(%
{\bf r}); \\
\chi _{1} &=&U_{11}\int d^{3}r\left| \phi _{1}({\bf r})\right| ^{4}; \\
\chi _{2} &=&U_{22}\int d^{3}r\left| \phi _{2}({\bf r})\right| ^{4}; \\
\chi _{12} &=&U_{12}\int d^{3}r\left| \phi _{1}({\bf r})\right| ^{2}\left|
\phi _{2}({\bf r})\right| ^{2}.
\end{eqnarray}

For $\Omega (t)=0$, the eigenstates of Eq. (\ref{mode}) are simply the
number states $\left| n_{1},n_{2}\right\rangle _{F}$ such that $\hat{N}%
_{i}\left| n_{1},n_{2}\right\rangle _{F}=n_{i}\left|
n_{1},n_{2}\right\rangle _{F}$ where $\hat{N}_{i}=a_{i}^{\dagger }a_{i}$.
One may note that Eq. (\ref{mode}) with $\chi _{12}=0$ also describes
tunnelling between two condensates in a double well potential\cite{milburn}.

\subsection{Angular momentum representation}

Equation (\ref{mode}) may be expressed in a more convenient form by taking
advantage of the mapping between the algebra for two independent harmonic
oscillators and the algebra for angular momentum \cite{sakurai}. The mapping
between the two algebras is achieved by making the following definitions 
\end{mathletters}
\begin{mathletters}
\begin{eqnarray}
J_{+} &=&a_{1}^{\dagger }a_{2}; \\
J_{-} &=&a_{2}^{\dagger }a_{1}; \\
J_{z} &=&\frac{1}{2}\left( a_{1}^{\dagger }a_{1}-a_{2}^{\dagger
}a_{2}\right) ;
\end{eqnarray}
By noting that the x and y components of the angular momentum are given by
the operators $J_{x}=\frac{1}{2}(J_{+}+J_{-})$ and $J_{y}=\frac{1}{2i}%
(J_{+}-J_{-})$, it follows that 
\end{mathletters}
\begin{equation}
{\bf J}^{2}=\frac{\hat{N}}{2}\left( \frac{\hat{N}}{2}+1\right) ;
\end{equation}
where $\hat{N}=\hat{N}_{1}+\hat{N}_{2}$ is the total number operator which
commutes with the two-mode Hamiltonian. Thus ${\bf J}^{2}$ is a constant of
motion with eigenvalues $j(j+1)$ and $j=N/2.$ Consequently, for a state with
definite $N$, Eq. (\ref{mode}) has the form \cite{collett}, 
\begin{equation}
\hat{H}=\Delta \omega J_{z}+\chi _{+}J_{z}^{2}+\frac{1}{2}\bar{\Omega}^{\ast
}(t)J_{+}+\frac{1}{2}\bar{\Omega}(t)J_{-}+\frac{1}{2}\left( \chi _{1}+\chi
_{2}+2\chi _{12}\right) j^{2}+\frac{1}{2}\left( 3\omega _{1}+3\omega
_{2}-\chi _{1}-\chi _{2}\right) j;  \label{Hjm}
\end{equation}
where 
\begin{mathletters}
\begin{eqnarray}
\Delta \omega &=&\delta +\frac{3}{2}\left( \omega _{1}-\omega _{2}\right)
+\chi _{-}(2j-1); \\
\chi _{+} &=&\frac{1}{2}\left( \chi _{1}+\chi _{2}-2\chi _{12}\right) ; \\
\chi _{-} &=&\frac{1}{2}\left( \chi _{1}-\chi _{2}\right) ;
\end{eqnarray}
In writing Eq. (\ref{Hjm}) we have made the replacement $\hat{N}\rightarrow
2j.$ This limits Eq. (\ref{Hjm}) to the subspace of states with the same
total number of atoms. However, in the following section, we are only
interested in calculating the expectation values of the operators $J_{z}$
and $J_{\pm }$ after some time $t$ so that the replacement $\hat{N}%
\rightarrow 2j$ does not affect any of our results even when we choose an
initial state that is a superposition of number states. Therefore, we can
drop the last two terms Eq. (\ref{Hjm}) so that our effective Hamiltonian is
given by 
\end{mathletters}
\begin{equation}
\hat{H}=\Delta \omega J_{z}+\chi _{+}J_{z}^{2}+\frac{1}{2}\bar{\Omega}^{\ast
}(t)J_{+}+\frac{1}{2}\bar{\Omega}(t)J_{-}.  \label{Hjm2}
\end{equation}

For $\Omega (t)=0$, the eigenstates of Eq. (\ref{Hjm}) are simply the
eigenstates of $J_{z}$, $\left| j,m\right\rangle $ with $m=-j,...,j.$ By
noting that $n_{1}=j+m$ and $n_{2}=j-m$, it follows that $\left|
j,m\right\rangle =\left| j+m,j-m\right\rangle _{F}.$ In order to avoid
confusion, we use the subscript $F$ ($F$ as in Fock) on the kets for the
number state basis in order to distinguish them from the angular momentum
kets. When $\bar{\Omega}(t)=0,$ the condensate ground state is simply the
lowest energy eigenstate of Eq. (\ref{Hjm}) for a fixed total number of
atoms. This corresponds to the $\left| j,m\right\rangle $ state with $m=%
\mathop{\rm integer}%
\left( -\frac{\Delta \omega }{2\chi _{+}}\right) $ for $\left| 
\mathop{\rm integer}%
\left( -\frac{\Delta \omega }{2\chi _{+}}\right) \right| <j$ and $m=\pm j$
otherwise. Here, $%
\mathop{\rm integer}%
()$ denotes the integer part of the number in parentheses. However, we
assume that $-\frac{\Delta \omega }{2\chi _{+}}<-j$ so that the ground
state, $\left| j,-j\right\rangle =\left| 0,N\right\rangle _{F}$, is fully
polarized. When the resonance condition $\delta +\frac{3}{2}\left( \omega
_{1}-\omega _{2}\right) =0$ is satisfied, $\left| j,-j\right\rangle $ will
be the ground state when {\it (i) }$\chi _{2}<\chi _{12}$ and $\chi _{+}>0$
or {\it (ii) }$\chi _{2}>\chi _{12}$ and $\chi _{+}<0.$

One might also consider a coherent state in the number state basis given by 
\begin{equation}
\left| 0,\alpha _{2}\right) =e^{-|\alpha _{2}|^{2}/2}\sum_{n_{2}=0}^{\infty }%
\frac{\alpha _{2}^{n_{2}}}{\sqrt{n_{2}!}}\left| 0,n_{2}\right\rangle
_{F}=e^{-|\alpha _{2}|^{2}/2}\sum_{n_{2}=0}^{\infty }\frac{\alpha
_{2}^{n_{2}}}{\sqrt{n_{2}!}}\delta _{n_{2},2j^{\prime }}\left| j^{\prime
},-j^{\prime }\right\rangle ,
\end{equation}
as a variational wave-function that minimizes $\hat{H}$ with $\alpha _{2}=%
\sqrt{N}e^{i\varphi }.$ One may show using Eq. (\ref{Hjm}) that $\delta
E=\left\langle j,-j\right| \hat{H}\left| j,-j\right\rangle -\left( 0,\alpha
_{2}\right| \hat{H}\left| 0,\alpha _{2}\right) =-\chi _{2}N/2$. Notice that $%
\chi _{2}=U_{22}/\left[ (2\pi )^{3/2}a_{ho,2}^{3}\right] \sim U_{22}/V$
where $V$ is the volume of the trap. Therefore $\delta E\sim U_{22}\left(
N/V\right) $ which is an intensive quantity. Since $\left\langle j,-j\right| 
\hat{H}\left| j,-j\right\rangle $ is extensive, $\delta E$ is negligible in
the thermodynamic limit. Therefore, the coherent state represents a good
variational wavefunction for the the ground state energy in the
thermodynamic limit.

The primary advantage of the angular momentum representation is that the
dynamics of the condensate can be understood in terms of a spin vector on a
Bloch sphere. For strong external pulses of duration $t_{p}$ such that $%
\int_{0}^{t_{p}}dt\left| \bar{\Omega}(t)\right| \gg \left| \Delta \omega
\right| t_{p},\left| \chi _{+}\right| t_{p},$ the time evolution operator, $%
U=\hat{T}e^{-i\int \hat{H}dt}$, is simply a rotation operator in spin-space 
\begin{mathletters}
\begin{equation}
R(\theta ,\phi )=\exp [-i\theta \left( J_{x}\sin \phi -J_{y}\cos \phi
\right) ],  \label{rot}
\end{equation}
where $\theta \sin \phi =\int_{0}^{t_{p}}dt%
\mathop{\rm Re}%
\left( \bar{\Omega}(t)\right) $ and $-\theta \cos \phi =\int_{0}^{t_{p}}dt%
\mathop{\rm Im}%
\left( \bar{\Omega}(t)\right) $. We have neglected the time ordering
operator, $\hat{T}$ in Eq. (\ref{rot}). This is justified if $\bar{\Omega}(t)
$ is a square pulse so that the Hamiltonian commutes with itself at
different times in the interval $0\leq t\leq t_{p}$. Equation (\ref{rot}) is
a rotation in spin-space through an angle $\theta $ about the rotation axis $%
{\bf n}=\left( \sin \phi ,-\cos \phi ,0\right) $. When there is no external
field present, the time evolution operator is simply 
\end{mathletters}
\begin{equation}
U_{o}(t)=e^{-i\left( \Delta \omega J_{z}+\chi _{+}J_{z}^{2}\right) t}
\label{free}
\end{equation}
which is diagonal in the $\left| j,m\right\rangle $ basis.

\section{Condensate Dynamics}

In this section, we consider the dynamics of the condensate subject to two
external pulses separated by a time interval $T$. For $t\leq 0$ we assume
that there have been no pulses applied and that the condensate is in the
ground state. We denote the two ground states, considered in the last
section, at $t=0$ by 
\begin{mathletters}
\begin{eqnarray}
\left| \Psi _{N}\right\rangle &=&\left| j,-j\right\rangle ; \\
\left| \Psi _{C}\right\rangle &=&\left| 0,\alpha _{2}\right) ;
\end{eqnarray}
These two states correspond to a spin vector that is pointing towards the
south pole of the Bloch sphere. For the number state, $\left| \Psi
_{N}\right\rangle $, the length of this vector in the $-z$ direction is $N/2$%
. The coherent state, $\left| \Psi _{C}\right\rangle ,$ also points in the $%
-z$ direction but with an {\it average} length of $N/2$ and an uncertainty
in the z-component of the length of $\Delta J_{z}=\frac{1}{2}\sqrt{N}.$

At time $t=0^{+},$ a pulse is applied that rotates the system about the
y-axis through an angle of $\frac{\pi }{2}.$ This pulse is described by the
rotation operator $R(\frac{\pi }{2},\pi ).$ It has the effect of
transferring half of the condensate population from state 2 into state 1 so
that after the pulse $\left\langle J_{z}\right\rangle =0$ with the spin
vector now pointing in the $+x$ direction. The condensate is then allowed to
evolve freely for a time $T$ according to Eq. (\ref{free}). After this
period of free evolution, a second $\frac{\pi }{2}$-pulse is applied, again
given by the rotation $R(\frac{\pi }{2},\pi ).$ After the second pulse, the
population difference, $\left\langle \hat{N}_{1}-\hat{N}_{2}\right\rangle
=2\left\langle J_{z}(T)\right\rangle $ as a function of $T$ is measured.

For $T=0$, the effect of the two pulses would just be a spin-flip so that
the spin vector would now be pointing in the $+z$ direction. Because the
state of the system following the first pulse (for both $\left| \Psi
_{N}\right\rangle $ and $\left| \Psi _{C}\right\rangle $) is a superposition
of $\left| j,m\right\rangle $ states for all $m$ values, the free evolution
due to Eq. (\ref{free}) causes the spin vector to diffuse in the equatorial
plane as the different $\left| j,m\right\rangle $ states get out of phase
with each other. Because the $m$ are discrete integers, the $\left|
j,m\right\rangle $ states can re-phase leading to a revival of the spin
vector. This dephasing and rephasing of the spin vector manifests itself as
collapse and revivals of the population difference following the second
pulse. However, the two initial states, $\left| \Psi _{N}\right\rangle $ and 
$\left| \Psi _{C}\right\rangle $, have very differenct collapse and revival
times owing to the fact that $\left| \Psi _{C}\right\rangle $ is a
superposition of states with different $j.$

The calculation of $\left\langle J_{z}(T)\right\rangle $ is straight forward
and we outline the calculation for the two intial states in the following
two subsections.

\subsection{Number state, $\left| \Psi _{N}\right\rangle .$}

The quantity we wish to calculate is 
\end{mathletters}
\begin{equation}
\left\langle J_{z}(T)\right\rangle _{N}=\left\langle \Psi _{N}\right| R(%
\frac{\pi }{2},\pi )^{\dagger }U_{o}(T)^{\dagger }R(\frac{\pi }{2},\pi
)^{\dagger }J_{z}R(\frac{\pi }{2},\pi )U_{o}(T)R(\frac{\pi }{2},\pi )\left|
\Psi _{N}\right\rangle .  \label{diff}
\end{equation}
First we note that 
\begin{equation}
R(\beta ,\pi )^{\dagger }J_{z}R(\beta ,\pi )=\cos \beta J_{z}-\sin \beta
J_{x}
\end{equation}
so that $R(\frac{\pi }{2},\pi )^{\dagger }J_{z}R(\frac{\pi }{2},\pi )=-%
\mathop{\rm Re}%
\left\{ J_{+}\right\} .$ Consequently, Eq. (\ref{diff}) reduces to 
\begin{equation}
\left\langle J_{z}(T)\right\rangle _{N}=-%
\mathop{\rm Re}%
\left\{ \left\langle \Psi _{N}\right| R(\frac{\pi }{2},\pi )^{\dagger
}U_{o}(T)^{\dagger }J_{+}U_{o}(T)R(\frac{\pi }{2},\pi )\left| \Psi
_{N}\right\rangle \right\} =-%
\mathop{\rm Re}%
\left\{ \left\langle J_{+}(T)\right\rangle _{N}\right\}
\end{equation}
The matrix elements of $R(\frac{\pi }{2},\pi )$ are easily calculated for
arbitrary $j$ \cite{sakurai}, so that the state of the system following the
first pulse and the free evolution period is simply, 
\begin{equation}
U_{o}(T)R(\frac{\pi }{2},\pi )\left| \Psi _{N}\right\rangle =\sum_{m=-j}^{j}%
\frac{(-1)^{m+j}}{2^{j}}\sqrt{\frac{(2j)!}{(j-m)!(j+m)!}}\exp \left[
-i\left( \Delta \omega m+\chi _{+}m^{2}\right) \right] \left|
j,m\right\rangle .
\end{equation}
Finally, one obtains 
\begin{mathletters}
\begin{equation}
\left\langle J_{+}(T)\right\rangle _{N}=-je^{i\Delta \omega T}\cos
^{2j-1}\left( \chi _{+}T\right) ;  \label{J+1N}
\end{equation}
so that the population difference, $\left\langle \hat{N}_{1}-\hat{N}%
_{2}\right\rangle _{N}=2\left\langle J_{z}(T)\right\rangle _{N}$ is then 
\end{mathletters}
\begin{equation}
\left\langle \hat{N}_{1}-\hat{N}_{2}\right\rangle _{N}=N\cos \left( \left(
\delta +\frac{3}{2}\left( \omega _{1}-\omega _{2}\right) +\chi
_{-}(N-1)\right) T\right) \cos ^{N-1}\left( \chi _{+}T\right) .
\label{fock-diff}
\end{equation}

\subsection{Coherent State, $\left| \Psi _{C}\right\rangle .$}

The calculation of $\left\langle J_{z}(T)\right\rangle _{C}$ is similar to
that of $\left\langle J_{z}(T)\right\rangle _{N},$the main difference being
an average over a Poissonian distribution of number states. As before, we
only need to calculate the expectation value of $J_{+}$ following the free
evolution period, 
\begin{equation}
\left\langle J_{z}(T)\right\rangle _{C}=-%
\mathop{\rm Re}%
\left\{ \left\langle \Psi _{C}\right| R(\frac{\pi }{2},\pi )^{\dagger
}U_{o}(T)^{\dagger }J_{+}U_{o}(T)R(\frac{\pi }{2},\pi )\left| \Psi
_{C}\right\rangle \right\} =-%
\mathop{\rm Re}%
\left\{ \left\langle J_{+}(T)\right\rangle _{C}\right\} 
\end{equation}
Following the first pulse and the free evolution period, the state of the
system is 
\begin{eqnarray}
U_{o}(T)R(\frac{\pi }{2},\pi )\left| \Psi _{C}\right\rangle  &=&e^{-|\alpha
_{2}|^{2}/2}\sum_{n_{2}=0}^{\infty }\frac{\alpha _{2}^{n_{2}}}{\sqrt{n_{2}!}}%
\delta _{n_{2},2j}  \label{coh} \\
&&\times \left( \sum_{m=-j}^{j}\frac{(-1)^{m+j}}{2^{j}}\sqrt{\frac{(2j)!}{%
(j-m)!(j+m)!}}\exp \left[ -i\left( \Delta \omega m+\chi _{+}m^{2}\right) %
\right] \left| j,m\right\rangle \right) ;  \nonumber
\end{eqnarray}
Again, it should be emphasized that Eq. (\ref{coh}) is only valid for
calculating matrix elements for operators that are diagonal in $j,$ which
include all angular momentum operators$.$ The evaluation of $\left\langle
J_{+}(T)\right\rangle _{C}$ can be done in two steps, 
\begin{equation}
\left\langle J_{+}(T)\right\rangle _{C}=e^{-|\alpha
_{2}|^{2}}\sum_{n_{2}=0}^{\infty }\frac{\left| \alpha _{2}\right| ^{2n_{2}}}{%
n_{2}!}\delta _{n_{2},2j}\left( -je^{i\Delta \omega T}\cos ^{2j-1}\left(
\chi _{+}T\right) \right) ,
\end{equation}
where the term in parantheses is the same as Eq. (\ref{J+1N}). The final
step is just an averaging over a Poissonian distribution of the number of
atoms ($|\alpha _{2}|^{2}=N$), 
\begin{equation}
\left\langle J_{+}(T)\right\rangle _{C}=-\frac{N}{2}e^{i\left( \delta +\frac{%
3}{2}\left( \omega _{1}-\omega _{2}\right) \right) T}\exp \left[ N\left(
-1+\cos \left( \chi _{+}T\right) e^{i\chi _{-}T}\right) \right] .
\end{equation}
The population difference following the second pulse is then, 
\begin{eqnarray}
\left\langle \hat{N}_{1}-\hat{N}_{2}\right\rangle _{C} &=&N\exp \left[
N\left( -1+\cos \left( \chi _{+}T\right) \cos (\chi _{-}T)\right) \right] 
\label{coh-diff} \\
&&\times \cos \left( \left( \delta +\frac{3}{2}\left( \omega _{1}-\omega
_{2}\right) \right) T+N\cos \left( \chi _{+}T\right) \sin (\chi
_{-}T)\right) .  \nonumber
\end{eqnarray}

Equations (\ref{fock-diff}) and (\ref{coh-diff}) represent the central
results of this paper. Note that although the number of atoms, $N,$
appearing in the two equations has the same value, the meaning of $N$ is
different. In Eq. (\ref{fock-diff}), $N$ is the exact number of particles
while for Eq. (\ref{coh-diff}), $N$ is the {\it average} number of particles
for a superposition of number states.

\section{Collapse and revivals}

From Eqs. (\ref{fock-diff}) and (\ref{coh-diff}) one can see that the
population difference involves a rapidly oscillating part and an envelope
function that is responsible for the collapse and revival of the population
difference. For simplicity we assume that the external field is on resonance
so that $\delta +\frac{3}{2}\left( \omega _{1}-\omega _{2}\right) =0$ and
the population difference for the two cases simplify to 
\begin{eqnarray}
\left\langle \hat{N}_{1}-\hat{N}_{2}\right\rangle _{N} &=&N\cos \left( \chi
_{-}(N-1)T\right) \cos ^{N-1}\left( \chi _{+}T\right) ;  \label{popN} \\
\left\langle \hat{N}_{1}-\hat{N}_{2}\right\rangle _{C} &=&N\exp \left[
N\left( -1+\cos \left( \chi _{+}T\right) \cos (\chi _{-}T)\right) \right]
\cos \left( N\cos \left( \chi _{+}T\right) \sin (\chi _{-}T)\right) ;
\label{popC}
\end{eqnarray}
In the following sub-sections we consider several limiting cases.

\subsection{$\protect\chi _{-}=0$}

The simplest nontrivial case to consider is $\chi _{1}=\chi _{2}\neq \chi
_{12}$ (this corresponds to Ref. \cite{wright2} where $\chi _{12}=0$) so
that $\left\langle \hat{N}_{1}-\hat{N}_{2}\right\rangle _{N}=N\cos
^{N-1}\left( \chi _{+}T\right) $ and $\left\langle \hat{N}_{1}-\hat{N}%
_{2}\right\rangle _{C}=N\exp \left[ N\left( -1+\cos (\chi _{+}T\right) )%
\right] .$ This case is shown in Figure 1. The population difference quickly
decays to zero for both cases as soon as $\cos (\chi _{+}T)$ deviates
significantly from $1.$ The collapse time may be estimated by making a
Gaussian approximation for small times. One finds then for $\chi _{+}T\ll 1$%
\begin{eqnarray}
\left\langle \hat{N}_{1}-\hat{N}_{2}\right\rangle _{N} &\approx
&Ne^{-(T/\tau _{N})^{2}}; \\
\left\langle \hat{N}_{1}-\hat{N}_{2}\right\rangle _{C} &\approx
&Ne^{-(T/\tau _{C})^{2}};
\end{eqnarray}
and the collapse times are $\tau _{N}=\chi _{+}^{-1}\sqrt{2/(N-1)}$ and $%
\tau _{C}=\chi _{+}^{-1}\sqrt{2/N}$ which are indistinguishable for $N\gg 1$%
. The variance in $J_{z}$ following the first pulse is given by 
\begin{equation}
\Delta J_{z}=\sqrt{\left\langle \Psi _{i}\right| R(\frac{\pi }{2},\pi
)J_{z}^{2}R(\frac{\pi }{2},\pi )\left| \Psi _{i}\right\rangle }=\sqrt{j/2}=%
\sqrt{N}/2
\end{equation}
for $i=N,C$. The collapse times can be expressed as $\tau _{N}=$ $\tau
_{C}=1/\left( \sqrt{2}\chi _{+}\Delta J_{z}\right) $ which shows that the
collapses are attributable to the dephasing of the different $J_{z}$ states
due to the $\chi _{+}J_{z}^{2}$ term in the Hamilitonian.

The revival times are quite different for the two states. For the number
state, the revivals occur whenever $T_{N}=n\pi /\chi _{+}$ where $n$ is an
integer. However, when the number of atoms is even, $N-1$ is odd and the
condensate will undergo {\it anti-revivals} when $T=(2n+1)\pi /\chi _{+}$ so
that $\left\langle \hat{N}_{1}-\hat{N}_{2}\right\rangle _{N}=-N$ at these
times. (Note that a similar affect was described in Ref. \cite{wright2} in
which the fringe visibility of the interference pattern could undergo a
revival with a $\pi $ phase shift when the number of atoms that had been 
{\it detected} in order to establish an interference pattern was even). In
contrast, the coherent state undergoes revivals at the times $T_{C}=2n\pi
/\chi _{+}$ which is twice the revival time of the number state. In
addition, $\left\langle \hat{N}_{1}-\hat{N}_{2}\right\rangle _{C}>0$
regardless of $N$. Therefore, there appear to be two key differences that
distinguish a number state from a coherent state for $\chi _{-}=0$: (i) The
occurrence of a negative population difference when $N$ is even and (ii)
revival times that are half the revival times of a coherent state.

Since the revivals are determined by the $\chi _{+}J_{z}^{2}$ in Eq. (\ref
{Hjm2}), there is a simple explanation for the factor of two difference in
the revival times. The state $U_{o}(T)R(\frac{\pi }{2},\pi )\left| \Psi
_{N}\right\rangle $ involves a superposition of $\left| j,m\right\rangle $
in which the $m$ values all differ by an integer. Consequently, the revivals
occur when the relative phase between all of the $\left| j,m\right\rangle $
states is an integer multiple of $2\pi .$ This corresponds to the condition $%
\left[ \left( m+1\right) ^{2}-m^{2}\right] \chi _{+}T=2n^{\prime }\pi +\phi $
for all $m$ and where $\phi $ is a global phase factor that is independent
of $m$ and $n^{\prime }$ is an integer. By taking $n^{\prime }=mn$ and $\phi
=\chi _{+}T$, one sees that the revivals occur at integer multiples of the
time $\pi /\chi _{+}$. However, for the coherent state one has instead $%
U_{o}(T)R(\frac{\pi }{2},\pi )\left| \Psi _{C}\right\rangle $ which is a
superposition of states with different $j$ and $m$ values so that in this
case the values of $m$ in the superposition need only differ by a {\it %
half-integer}. Therefore, for the coherent state the condition for the
occurrence of a revival is $\left[ \left( m+1/2\right) ^{2}-m^{2}\right]
\chi _{+}T=2n^{\prime }\pi +\phi .$ Again taking $n^{\prime }=mn$ but with $%
\phi =\chi _{+}T/4,$ the revivals occur at integer multiples of $2\pi /\chi
_{+}$.

\subsection{$\protect\chi _{-}\neq 0$ and $\protect\chi _{-}N\gg 1$}

In this case Eqs. (\ref{popN}-\ref{popC}) consist of a rapid oscillations
modulated by a slowly varying envelope function which gives rise to the
collapse and revivals. As such, we only consider the behavior of the
envelope functions in this sub-section which is given by $f_{N}(T)$ and $%
f_{C}(T)$ for the number state and coherent state, respectively. The
envelope functions are given by 
\begin{mathletters}
\begin{eqnarray}
f_{N}(T) &=&\cos ^{N-1}\left( \chi _{+}T\right) , \\
f_{C}(T) &=&\exp \left[ N\left( -1+\cos \left( \chi _{+}T\right) \cos (\chi
_{-}T)\right) \right] .
\end{eqnarray}

The collapse and revival times for the number state are the same as what was
found in the previous subsection. The only difference is that there are no
antirevivals since $f_{N}(T)=-1$ simply corresponds to a $\pi $ phase shift
in the rapidly oscillating part of Eq. (\ref{popN}).

However, the behavior of the coherent state is quite different. One can see
that for short times, $f_{C}(T)\approx e^{-(T/\tau _{C})^{2}}$ where the
collapse time is $\tau _{C}=\sqrt{\frac{2}{N(\chi _{+}^{2}+\chi _{-}^{2})}}.$
Therefore, increasing $\chi _{-}$ decreases the collapse time. This is
attributable to the $\chi _{-}(2j-1)J_{z}$ term in the Hamiltonian which
causes the states with different $j$ but the same $m$ to get out of phase
with each other in a time $\sim 1/(\chi _{-}\Delta N)=1/(\chi _{-}\sqrt{N}).$
The reduction in the collapse time is illustrated in Figure 2 for the case $%
\chi _{-}/\chi _{+}=2.$

Revivals occur when $\cos \left( \chi _{+}T\right) \cos (\chi _{-}T)=1$
which can only be satisfied if $\chi _{-}/\chi _{+}=p/q$ where $p$ and $q$
are integers. When this condition is satisfied, the revivals occur at times $%
T_{C}=n\pi /\chi _{+}$ where $n$ is a positive integer {\it and} $\left(
p/q\right) n$ is an integer such that if $n$ is odd (even) then $\left(
p/q\right) n$ is also odd (even). When $\chi _{-}/\chi _{+}$ is irrational,
there are no revivals and even for rational values of $\chi _{-}/\chi _{+},$
the period between revivals can differ significantly from the revival period
for the number state, $\pi /\chi _{+}.$ For example, if $\chi _{-}/\chi
_{+}=1/4$ then the first revival will occur at $8\pi /\chi _{+}$ for the
coherent state. In Figure 3, the revivals are shown for the ratio $\chi
_{-}/\chi _{+}=1/3.$ One can see that the number state has revivals at all
integer multiples of $\pi /\chi _{+}$ while the first revival for the
coherent state occurs at $3\pi /\chi _{+}.$

\section{Discussion}

In the previous section it was shown that the population difference in a
Ramsey fringe experiment can exhibit collapses and revivals with times that
can be very different for a number state and a coherent state. For a number
state, we will, in general, be ignorant of the exact number of atoms so that
we can only say that the state $\left| 0,N\right\rangle $ occurs with some
probability $p(N).$ Consequently, an ensemble average over many different
experimental runs will lead to statistical fluctuations in the number of
atoms even though for the number state, the number of atoms in any given
experiment would be precisely defined. Suppose $p(N)$ obeys a Poisson
distribution, will there be any difference now between the number states and
the coherent state? There is in fact a difference since the quantum
fluctuations of the coherent state are present in each experimental run
whereas the statistical fluctuations of the number states only become
manifest after averaging over many such runs. As an example, consider the
case of the anti-revivals for the number state, $\left\langle \hat{N}_{1}-%
\hat{N}_{2}\right\rangle _{N}=-N$, which can occur when $\chi _{1}=\chi
_{2}\neq \chi _{12}.$ In this case, one might observe such anti-revivals in
any {\it single} experiment with a number state but such anti-revivals would
never be observed in any experiment for a coherent state since it is the
averaging over the Poisson number distribution which prevents the
anti-revivals. Even though the anti-revivals may be observed in each
experimental run for the number states, the average over many experimental
runs, $\sum_{N}p(N)\left\langle \hat{N}_{1}-\hat{N}_{2}\right\rangle
_{N}=\left\langle \hat{N}_{1}-\hat{N}_{2}\right\rangle _{C}$, will not
exhibit the anti-revivals since they get averaged out.

One may also consider initial states of the condensate given by $\left| \Psi
\right\rangle =\sum_{n}c_{n}\left| 0,n\right\rangle _{F}$ where $c_{n}$ is
sharply peaked around $n=N$ (such as $c_{n}\propto e^{-(n-N)/4\sigma ^{2}}$%
). This state satisfies Eq. (\ref{SBGS}) for finite $N.$ However, $\left|
\Psi \right\rangle $ will exhibit qualitatively similar behavior to the
coherent state \cite{imamoglu} since the critical difference in the results
of the Ramsey fringe experiment lie in the quantum fluctuations in the
particle number of the coherent state.

Finally, we estimate the order of magnitude of the collapse and revival
times. The fundamental time scale which determines the collapses and
revivals is $\hbar /\chi _{+}$ (where we now explicitly include that factors
of $\hbar $). If the trapping potentials for the two components are the
same, $\omega _{1}=\omega _{2}=\omega ,$ then, $\phi _{1}(r)=\phi
_{2}(r)=\phi (r)$ and $\int \left| \phi (r)\right| ^{4}d^{3}r=1/\left[ (2\pi
)^{3/2}a_{ho}^{3}\right] $ where $a_{ho}=\sqrt{\hbar /m\omega }.$ If we
assume a relatively weak trap, $a_{ho}=5\mu m$ and if we take $U_{ij}=\frac{%
4\pi \hbar ^{2}a_{ij}}{m}\sim \frac{4\pi \hbar ^{2}a}{m}$ with $a=5nm$ as an
estimate for $^{87}Rb$, one finds $\chi _{i}/\hbar \sim \left( \frac{4\pi
\hbar a}{m}\right) /\left[ (2\pi )^{3/2}a_{ho}^{3}\right] =0.023s^{-1}.$
Therefore the collapse times will be on the order of $\left( \chi _{i}/\hbar
\right) ^{-1}\sqrt{1/N}\sim 1s$ for $N=1000$ and the revival times will be
on the order of $\pi \left( \chi _{i}/\hbar \right) ^{-1}\sim 100s.$ Note
that since the scattering lengths are nearly equal for the two hyperfine
states of the $^{87}Rb$ condensate \cite{Rb}, $\chi _{\pm }\approx 0$ for $%
\omega _{1}=\omega _{2}.$ However, this may be overcome by either
manipulating one of the scattering lengths using a Feshbach resonance\cite
{fesh} or by changing the trapping potentials so that $\phi _{1}(r)\neq \phi
_{2}(r)$.

As mentioned, a Ramsey fringe experiment has been performed at JILA \cite
{relphase}. However, this experiment was performed with a relatively large
condensate ($N=5\times 10^{5}$) such that $Na_{ij}/a_{ho,i}\gg 1$ \cite{mode}%
. The theoretical analysis of the experiment \cite{relphase} \cite{ramsey2}
was based on the Gross-Pitaevskii equation which is a mean-field equation
for the order parameter and, as such, already assumes a state of broken
symmetry in the condensate. In addition, the duration of this experiment
(i.e. the period between pulses, $T$) was too short too observe the phase
diffusion due to the quantum dynamics of the condensate (see footnote [23]
in Ref. \cite{relphase}).

\section{Conclusion}

In this paper we have shown that Ramsey's separated oscillatory field
technique applied to a small atomic Bose-Einstein condensate exhibits
collapse and revivals in the population difference between the two internal
states of the condensate. The collapse and revival times depend on the
strength of the two-body interactions and the initial state of the
condensate so that one may potentially distinguish between a condensate
state that is a number state or a coherent state. Since absorptive and
dispersive imaging of atomic BEC measure the density of atoms in the
condensate, the Ramsey fringe experiment proposed here should be easier to
perform than an experiment which tries to directly observe the collapse and
revivals in the order parameter.

\end{mathletters}

\end{document}